# Real-time inclinometer using accelerometer MEMS

D. Hanto[1], B. Widiyatmoko[1], B. Hermanto[1], P. Puranto[1], and L.T. Handoko[1,2]

[1] Research Center for Physics, Indonesian Institute of Sciences
Kompleks Puspiptek Serpong, Tangerang 15310, Indonesia
[2] Department of Physics, University of Indonesia
Kampus Depok, Depok 16424, Indonesia

*Abstract*— **A preliminary design of inclinometer for real-time monitoring system of soil displacement is proposed. The system is developed using accelerometer sensor with microelectromechanical system (MEMS) device. The main apparatus consists of a single MEMS sensor attached to a solid pipe and stucked pependicularly far away below the soil surface. The system utilizes small fractions of electrical signals from MEMS sensor induced by the pipe inclination due to soil displacements below the surface. It is argued that the system is accurate enough to detect soil displacements responsible for landslides, and then realizes a simple and low cost landslide early warning system.**

*Keywords – landslides; inclinometer; MEMS accelerometer, sensor*

## I. Introduction

Ground movement which induce landslides or landslip are a common terminology to speak for a process of the movement of earth material mass from one point to anothers. This geological phenomenon is mainly driven by gravitation force. However, there are also another contributing factors caused by human or nature which could destroy the slope stability, such as groundwater pressure, the loss of soil structure, erotion, earthquake and so on [1,2].

The landslide hazard analysis and mapping are crucial to reduce potential catastrophic loss, and on the other hand to guide sustainable land use planning as well. The above-mentioned factors are usually categorized into geomorphology, geology, land use / cover and hydrogeology [3]. Fortunately, these uncountable factors affecting the landslide always accumulate as a phenomenon of soild displacements below the surface. This fact leads to some efforts to detect the underground soil displacements as a useful tool to predict and map the incoming landslides.

One of the widely adopted technique to identify and map the soil displacements in high accuracy is the Geographic Information System (GIS) [4,5]. Remote sensing techniques are also highly employed for landslide hazard assessment and analysis. Comparing the aerial photographs and satellite imagery through a period of time can provide some information on landslide characteristics, like distribution and classification, and factors like slope, lithology, and land use/land cover to be used to help predict future events [6].

Although its superiority, GIS and remote sensing techniques require advanced technology and expertises that might not be affordable in most regions especially in the developing countries. This point is crucial in term of landslide early warning system (LEWS). Because the analysis could spread over wide area with densed population. Some research groups have developed integrated LEWS, however most of them are focused on quantifying the hydrogeological factors like erotion, river flows as done by the Sensor-based LEWS (SLEWS) in Germany [7]. On the other hand, there is an approach relying on the extensometer and automatic rain gauge to predict the landslides in certain area [8,9].

The existing approaches so far rely on some quantitative modeling to relate the external factors and the probability of landslide occurancy. Despite the method seems complete and covers as many as possible all factors, the information is not useful as expected. An ideal LEWS should satisfy some principal requirements :

- Accurate : LEWS should be able to provide the incoming landslides as soon as possible with controllable errors.

- Low cost and easy to implement : covering as wide as possible area is important to improve the accuracy and reduce the potential disasters.

- Low cost and less maintainance : LEWS should work around the clock in 24/7 base. So the whole system should have high durability against heavy environments.

Considering these requirements, the existing LEWS' are still far from the expectation. The modeling based LEWS requires pra-observations and mathematical modeling which should be calibrated thereafter. On the other hand, the prediction is always under question and within significant error.

This paper presents a novel approach to overcome such problems. Rather than developing a system at an early step of external factors causing the landslides, we propose a straightforward approach by monitoring the underground soil displacements in a real-time base using accelerometer MEMS device. However, this paper is focused only on the preliminary design using a single MEMS sensor.

The paper is organized as follows. After this introduction, brief introduction on MEMS accelerometer and the design of LEWS using MEMS sensor are given in Sec. II. Before sumaryzing the paper, the expected level of accuracy is discussed briefly.

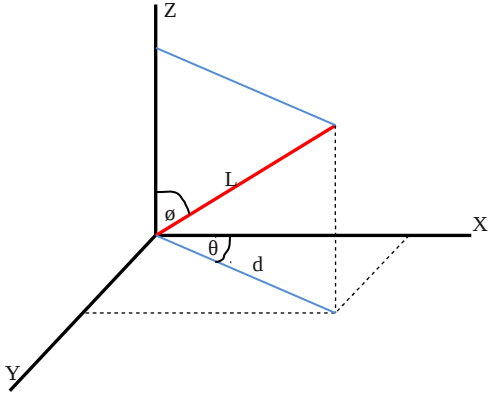

Figure 1. The projection of inclinometer on 2-dimensional *X-Y* plane.

## II. Inclinometer using MEMS sensor

First of all, let us briefly review the mechanism of MEMS-based accelerator sensor. In general, accelerometer is intended to measure proper acceleration, that is the acceleration experienced relative to freefall. There are both single- and multi-axes models which are able to detect magnitude and direction of the acceleration as a vector quantity. The quantity can then be used to determine orientation, acceleration, vibration shock and falling.

The basic principle of any accelerometers is measuring the displacement of the damped mass attached at a spring. When the accelerometer experiences an acceleration, the mass is displaced to the point that the spring is able to accelerate the mass at the same rate as the casing.

### A. MEMS accelerometer ADXL330

In addition to the conventional accelerometer, the MEMS acceletometer contains a tiny cantilever beam with a proof / seismic mass [10]. The sensor sends an analog signal as the proof mass deflects from its neutral position under the influence of external acceleration.

In this paper, MEMS accelerometer made by Analog Devices Inc., ADXL330, is particularly used [11]. This model already deploys multi-axes sensor in three-dimensional space. This provides a great advantage for the current interest since one does not need to set up at least two small MEMS acceletometers perpendicularly to obtain two dimensional orientation.

Using ADXL330, it is able to measure the inclinations of object relative to 2-dimensional plane simultaneously. The devise is also small in size, 4 x 4 x 1.45 (mm) and consumes low power and integrates three-axes sensor in a single circuit. It also can measure both dynamic and static accelerations with enough accuracy, that is within -3g and +3g, and sensitivity equal to 320 mV / 1 gr [11].

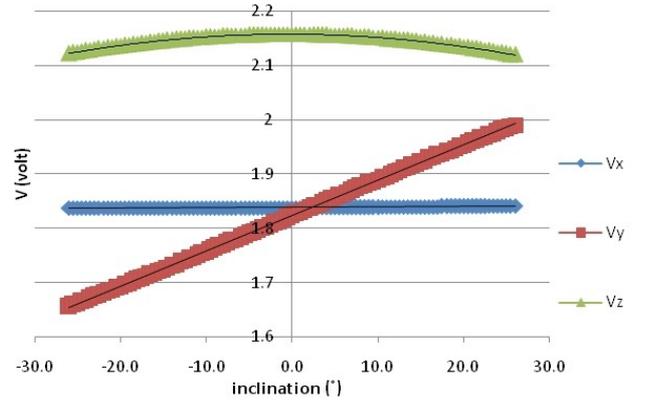

Figure 2. The output s of ($V_X$, $V_Y$, $V_Z$) represent the inclination of sensor around *X*-axis.

### B. The design of real-time inclinometer

Concerning the characteristics of MEMS accelerometer as ADXL330 above, one can consier to utilize it as a monitoring device to detect the change of inclination.

The idea is depicted in the schematic figure in Fig. 1. The sensor is attached at one edge of a solid pipe which is stucked perpendicularly inside the ground. The pipe length *L* is adjustable according to the depth of moving soil under observation.

The reference value is defined as the initial value at a completely perpendicular position. Therefore, discrepancies of the outputs ($V_X$, $V_Y$, $V_Z$) from each sensor can be converted as the magnitude of inclination in certain direction after proper signal proceesing and calibration. Because the difference of $V_x$ ($V_Y$) reflects the inclination around *Y* (*X*) -axis. For illustration, the real outputs from each orientation are shown in Figs. 2 and 3.

## III. Summary and discussion

A preliminary design of real-time inclinometer using a single MEMS accelerometer has been explained. It has been shown that the discrepancies of outputs ($V_X$, $V_Y$, $V_Z$) are precise enough to monitor the inclinations of solid pipe where the sensor is attached.

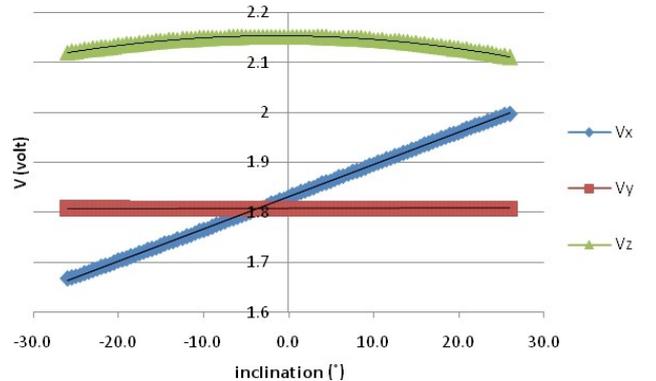

Figure 3. The output s of ($V_X$, $V_Y$, $V_Z$) represent the inclination of sensor around *Y*-axis.

Further works are still in progress to realize a comprehensive LEWS, that is :

- Mathematical algorithm for data acquisition, data processing and final analysis to monitor in real-time the magnitude, depth and location of ground displacements which might lead to landslides.
- Determine the most efficient combinations of several sensors to obtain as many and accurate as possible the ground displacements below the surface.
- Develop an integrated system consisting of multi sensors, software-based signal processing, data processing, wireless data transfer and web-based system information for public access.

The final results will be reported elsewhere in the near future.

ACKNOWLEDGMENT

The work is supported by Riset Kompetitif LIPI 2010 under contract number : 11.04/SK/KPPI/II/2010.

REFERENCES

[1] D. J. Easterbrook, "Surface Processes and Landforms. Upper Saddle River", Prentice-Hall, 1999.
[2] R.L. Schuster, R.J. Krizek, "Landslides: Analysis and Control", National Academy of Sciences, 1978.
[3] A. Clerici, S. Perego, C. Tellini, P.Vescavi, "A procedure for landslide susceptibility zonation by the conditional analysis method", Geomorphology 48, pp. 349-364, 2002.
[4] N. De La Ville, A.C. Diaz, D. Ramirez, "Remote sensing and GIS technologies as tools to support sustainable management of areas devastated by landslides", Environment, Development, and Sustainability 4, pp. 221-229, 2002.
[5] A. Fabbri, C. Chung, A. Cendrero, J. Remondo, "Is prediction of future landslides possible with a GIS?, Natural Hazards 30, pp. 487-499, 2003.
[6] G. Metternicht, L. Hurni, R. Gogu, "Remote sensing of landslides: An analysis of the potential contribution to geo-spatial systems for hazard assessment in mountainous environments", Remote Sensing of Environment 98, pp. 284-303, 2005.
[7] SLEWS Project, http://www.slews.de.
[8] T.F. Fathani, D. Karnawati, K. Sassa, H. Fukuoka, K. Honda, "Development of landslide monitoring and early warning system in Indonesia", Proceeding of the 1st World Landslide Forum, ICL UNESCO, Tokyo Japan, pp. 195-198, 2008.
[9] M. Rossi, S. Peruccacci, F. Guzzetti, M. Cardinali, P. Reichenbach, F. Ardizzone, P. Salvati, M.T. Brunetti, A. Mondini, A. Corazza, F. Leone, G. Tonelli, "Italian landslide early warning system", Geophysical Research Abstracts 11, EGU2009-2558, 2009.
[10] J. W. Judy, "Microelectromechanical systems (MEMS): fabrication, design and applications", Smart Materials and Structures 10, pp. 1115, 2001.
[11] Analog Devices Inc., "ADXL330: Small, Low Power, 3-Axis ±3g iMEMS® Accelerometer", http://www.analog.com, accessed 20 June 2010.